*Research Article*

# An M-QAM Signal Modulation Recognition Algorithm in AWGN Channel

**Ahmed K. Ali**  **and Ergun Erçelebi**

*Department of Electric and Electronic Engineering, Gaziantep University, Gaziantep 27310, Turkey*

Correspondence should be addressed to Ahmed K. Ali; engineer28ahmed@gmail.com

Computing the distinct features from input data, before the classification, is a part of complexity to the methods of automatic modulation classification (AMC) which deals with modulation classification and is a pattern recognition problem. However, the algorithms that focus on multilevel quadrature amplitude modulation (M-QAM) which underneath different channel scenarios is well detailed. A search of the literature revealed that few studies were performed on the classification of high-order M-QAM modulation schemes such as 128-QAM, 256-QAM, 512-QAM, and 1024-QAM. This work focuses on the investigation of the powerful capability of the natural logarithmic properties and the possibility of extracting higher order cumulant's (HOC) features from input data received raw. The HOC signals were extracted under the additive white Gaussian noise (AWGN) channel with four effective parameters which were defined to distinguish the types of modulation from the set: 4-QAM~1024-QAM. This approach makes the classifier more intelligent and improves the success rate of classification. The simulation results manifest that a very good classification rate is achieved at a low SNR of 5 dB, which was performed under conditions of statistical noisy channel models. This shows the potential of the logarithmic classifier model for the application of M-QAM signal classification. furthermore, most results were promising and showed that the logarithmic classifier works well under both AWGN and different fading channels, as well as it can achieve a reliable recognition rate even at a lower signal-to-noise ratio (less than zero). It can be considered as an integrated automatic modulation classification (AMC) system in order to identify the higher order of M-QAM signals that has a unique logarithmic classifier to represent higher versatility. Hence, it has a superior performance in all previous works in automatic modulation identification systems.

## 1. Introduction

Efficacious information transmission can be realized very clearly in trendy communication systems, and the transmitted signals are typically modulated by using various modulation ways. Modulation recognition is an intermediate way that must usually be achieved before signal demodulation and information detection, and it represents the substantial feature in modern radio systems to give knowledge on modulation signals rather than signal demodulation and can be used in decoding both civilian and military applications such as cognitive radio, signal identification, menace assessment, spectrum senses, and management, which allows to more efficiently use the available spectrum and increase the speed of data transfer. Furthermore, the unknown signal classification is a decisive weapon in electronic warfare scenarios. The electronic support management system plays a paramount role as a source of information is required to conduct electronic counter repression, threat analysis, warning, and target acquisition.

Particular recognition to modulation is the identification of types of the transmitted signals that lie in noncooperative channel environment groups which are significant for following up the signal demodulation and data extraction, and this was considered as the major turn to automatic modulation classification (AMC) which became an attractive subject for researchers, yet there is a major challenging for engineers who deal with the design of software-defined radio systems (SDRSs) and transmission deviation. The implementation of sophisticated information services and systems to military applications under a congested electromagnetic spectrum is a major concern issue to communication engineers. Friendly signals must be safely transmitted and received, whilst enemy signals should be found and jammed [1].

The first base of AMC theory is provided in [2], the essential ideas proposed at Stanford University in domestic project and was documented since 1969, and nowadays, the AMC is extensively used and well known for communication engineers as a significant part in the internal design of intelligent radio systems [3, 4].

Currently, digital modulation identification algorithms can be split into two techniques: a maximum likelihood hypothesis method which is based on decision theory [5] and statistical pattern recognition which relies upon the feature extraction ideas [6, 7], in uncooperative channel environment. The algorithms of the pattern recognition technique are usually used in practical application, and that is due to the absence of prior knowledge from received modulation signals [8]. The discrimination method has particularly beneficial features for pattern recognition which are initially extracted from the vectors of data and the identification of the signal modulation mode that was completed upon the coverage between characteristic parameters and limits of the extracted features which was considered as the known mode of the modulation type [9].

In the last five years, a new modulation classification technique has been proposed at the same time with the evolution of the automatic digital signal modulation recognition algorithms research, and this new modulation offers a new trend of complicated signal sets and higher order modulation signals. The most commonly used features are high-order cumulants (HOC) and high-order moment (HOM) [10]. The HOC-based techniques show superior performance to classify the digital modulation signals at a low signal-to-noise ratio (SNR); these algorithms were documented in literature [11–13]. Nevertheless, they were not appropriate to classify higher-order M-QAM modulation formats. Constellation shape varies from one modulation signal type to another in order to be considered as a robust signature.

The constellation shape was used in the algorithms in references [14–16] to recognize various types of higher order modulation including 8-PSK and 16-QAM. The recognition of success rate probabilities is around 95%, but it still required more SNR which may reach 4 dB. However, the constellation shape can be classified into M-PSK and M-QAM, and at the same time, they are sensitive to several wireless channels that could seriously make confusions in work; these comprise frequency offset, phase rotation, and the application of raised cosine roll-off filters. Under this situation, the symbol rate and frequency offset setting must be more accurate during the periods of presentment, and at the same time, the most widely used features are a cyclic spectrum and cyclic frequency that could help multiple unwanted signals to be detected with each of temporal and frequency-based overlaps [17–20].

In 2012, Dobber and his coworkers [21] suggested two novel algorithms for the recognition of 4-QAM, 16-QAM, 64-QAM, and QAMV.29 modulation schemes; also other [22, 23] authors suggested to use higher order cyclic cumulants (CCs) which were obtained from received signals as features for modulation classification. Also, this had been put forward in [24] which presents a method to classify a number of mixed modulation schemes with an assortment of N-class problems, and this technique also achieved a notable performance at low ranges of SNR. Due to the development of algorithms in machine learning, the researchers had started to design classification algorithms based on some aspects of machine learning which improved the classification ability and gives further pros to the classifier in terms of distinguished types of modulation signals.

Some investigations which were achieved on the behaviors of hidden naive Bayes (HNB) [25] and on combination of naive Bayes (NB) and other types of classifiers to create a new classifier named multiple classifier [26] certified that such classifier types were highly effective compared to conventional classifiers in terms of a small number of training iterations.

Supervised learning technique was combined with a modified K-means algorithm that is based on four famous optimization algorithms. This modification presented a new generation of the AMC technique. Likewise, artificial neural networks (ANNs) with genetic algorithms were studied and considered by Norouz and his coworkers [27], and also support vector machine (SVM) classifier in [28, 29].

The ways based on ANN and SVM show a superior performance although there is less prior information of signal features. However, multiple training samples and a long training period are necessary to accomplish sufficient learning which increases the computational complexity and makes the processing time longer.

In the last five years, researches widely explored the new techniques in order to reduce the required SNR and make it more efficient of recognition capability through focusing on robust features and classifier designs. One of the weak points of the previous algorithms is that the nature of the decision-tree is that it requires fixed threshold values due to the features that had been proposed by the authors, and these features are highly sensitive to any changes in SNR that can make the threshold values be valid for small ranges of SNR above to 10 dB. However, there had been no work until now that concentrates on the recognition of higher order QAM signals in terms of the shape of feature distribution curves.

In fact, this assumption gives a good understanding of the behavior of systems and reflects their major trends; therefore, under the circumstance of a channel corrupted by AWGN, this paper derived a relationship among of the higher order cumulant as features and threshold levels, in order to evaluate the performance of M-QAM modulation recognition technique.

## 2. Mathematical Analyses

### 2.1. Logarithmic Calculation.
Below are some theoretical analyses that assumes two logarithmic functions denoted $f_1(x_1)$ and $f_2(x_2)$ carrying different variable "$x_1$, $x_2$" but have the same base value "$n$" which can be written as follows:

$$f_1(x_1) = \log_n[x_1],$$
$$f_1(x_2) = \log_n[x_2], \tag{1}$$

where the range of variables $x_1 \neq x_2$ and the base value is considered to be the same. The ratio between $f_1(x_1)$ and $f_2(x_2)$ can be expressed as

$$W(x_1, x_2) = \frac{\log_n[x_1]}{\log_n[x_2]} = \frac{\ln[x_1]/\ln(n)}{\ln[x_2]/\ln(n)}. \tag{2}$$

In general, the logarithmic equations can be expressed into another equivalent form, written as

$$\log_n[x_1] = \frac{\ln[x_1]}{\ln(n)}. \quad (3)$$

Since $\ln[x]$ is a log function, it can also be expressed as $\log_r[x]$, where "$r$" indicates the base value of natural logarithms. Therefore, also equation (2) can be calculated as

$$W(x_1, x_2) = \frac{\ln[x_1]/\ln(n)}{\ln[x_2]/\ln(n)}. \quad (4)$$

Due to the equality between numerator and denominator, the $\ln(n)$ part has been eliminated. This gives the result expressed as below:

$$W(x_1, x_2) = \frac{\ln[x_1]}{\ln[x_2]}, \quad (5)$$

where $x_2$ is the constant value "10" that makes $w(x_1, x_2)$ a constant as well. The logarithmic functions are defined based on the higher order cumulant. In Section 3, the feature distribution curves with logarithmic properties are obtained, and the plot depicts that the modulation scheme is less sensitive to the variation of SNR.

## 3. Parameter Extraction Based on Logarithmic Properties

*3.1. Parameter Extraction Based on HOC.* It is well notable that a random variable $y(t)$ under complex-stationary process always has a zero mean value. Therefore, the following transformation formula which is based on higher order moments can be accomplished:

$$\begin{aligned}
C_{11} &= \text{cum}(y(n)^*, y(n)^*) = M_{11}, \\
C_{22} &= \text{cum}(y(n)^*, y(n)^*) = M_{22} - M_{20}^2 - 2M_{11}^2, \\
C_{33} &= \text{cum}(y(n)^*, y(n)^*, y(n)^*) = M_{33} - 6M_{20}M_{31} \\
&\quad - 9M_{22}M_{11} + 18(M_{20})^2 M_{11} + 12M_{11}^3, \\
C_{44} &= \text{cum}(y(n)^*, y(n)^*, y(n)^*, y(n)^*) = M_{44} - M_{40}^2 \\
&\quad - 18M_{22}^2 - 54M_{20}^4 - 144M_{11}^4 - 432M_{20}^2 M_{11}^2 \\
&\quad + 12M_{40}M_{20}^2 + 192M_{31}M_{11}M_{20} + 144M_{22}M_{11}^2 \\
&\quad + 72M_{22}M_{20}^2,
\end{aligned} \quad (6)$$

where the sign $\text{cum}(..)$ represents the cumulant operation, while the superscript "$*$" represents the operation of complex conjugate [30].

*3.2. Improved Higher Order Cumulant Feature.* The logarithmic expression formulas in (7)–(10) below show the modified higher order cumulant in terms of logarithm. The pros of modification not only make these features insensitive to noise but also help classifiers to recognize the increase in the higher order modulation signals. The simulation condition tests of 10,000 signal realizations from {4~1024}-QAM each consisting signal length $N = 4096$ with a phase offset of "$\pi/6$" with the statistical average value are shown in Figures 1–4, the distribution curve of each feature with varying values of the SNR. Through the simulation results, it is clear that the feature distribution curve is not significantly affected by variation of noise. However, this modification provides a better improvement in the achieved classification activity:

$$f_a = \log_{10}[|C_{11}|], \quad (7)$$

$$f_b = \log_{10}[|C_{22}|], \quad (8)$$

$$f_c = \log_{10}[|C_{33}|], \quad (9)$$

$$f_d = \log_{10}[|C_{44}|]. \quad (10)$$

## 4. Details on Algorithm and Simulation Performance

The above four HOC parameters have been derived and modified by using the logarithmic function to get optimum quality to recognize the M-QAM signals in the AWGN channel. However, the basic procedure of the automatic modulation classification system is shown in Figure 5 [31], while the actual decision procedure of the algorithm that has been proposed is demonstrated in Figure 6, yet the logarithmic modulation recognition method is effective with nine sets of M-QAM signals. The feature distribution curves in Figures 1–4 show the extracted features of modulation formats. Feature extraction is a primary step in the pattern recognition method. Threshold settings can then be determined depending on the distribution conditions of the extracted features. The threshold levels have been determined based on the projection of two lines upward and downward with effective space to avoid misclassification. The lines are parallel to the SNR axis (within a certain range of SNR). The created cluster region is used to separate required class distribution samples (modulation format that required discrimination). The points are computed empirically on the $y$-axis (feature distribution level), and this criterion will give a highly effective recognition rate even with a minimum value of SNR.

Table 1 represents the feature threshold values which were set empirically. The Boolean equations (12)–(16) illustrated the logical decision process, while Figure 7 shows the block diagram of the M-QAM logarithmic recognition algorithm, which is posterior to easily implement an extension of the proposed method in real time.

Algorithm 1 formally describes the procedures for the feature extraction and the decision for modulation recognition. On the contrary, in the feature extracting procedure, the input is represented by the data row after modulated with M-QAM. However, $s_i = d_x$ signal passes through a noisy channel $s_i = d'_x$, and the first step in extracting is fetching $\{C_{11}, C_{22}, C_{33}, C_{44}\}$; after computing the cumulant vector of signals, the process of the logarithmic classifier is begun and tested if $\{f_{a_0} < f_x(\varphi_0)\}$, if so $\text{index}_1 = 1$ for the 4-QAM signal.

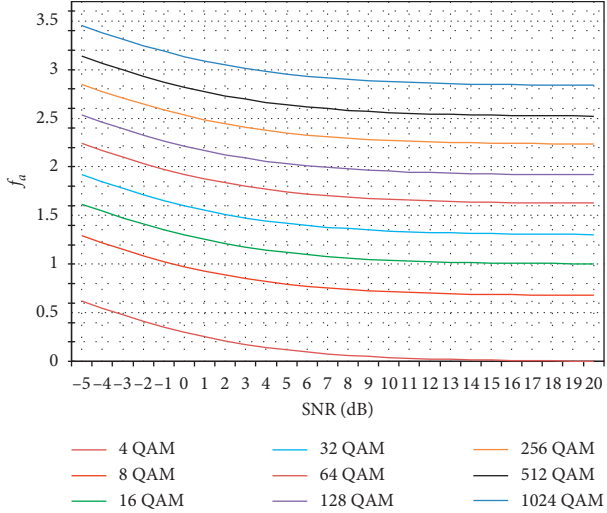

Figure 1: Feature "$f_a$" distribution curve vs SNR with the test of 10,000 signal realizations from {4~1024}-QAM each consisting signal length $N = 4096$ with a phase offset of "$\pi/6$."

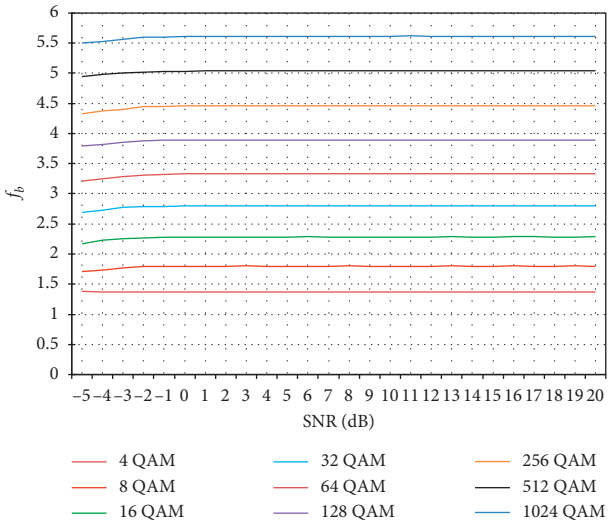

Figure 2: Feature "$f_b$" distribution curve vs SNR with the test of 10,000 signal realizations from {4~1024}-QAM each consisting signal length $N = 4096$ with a phase offset of "$\pi/6$."

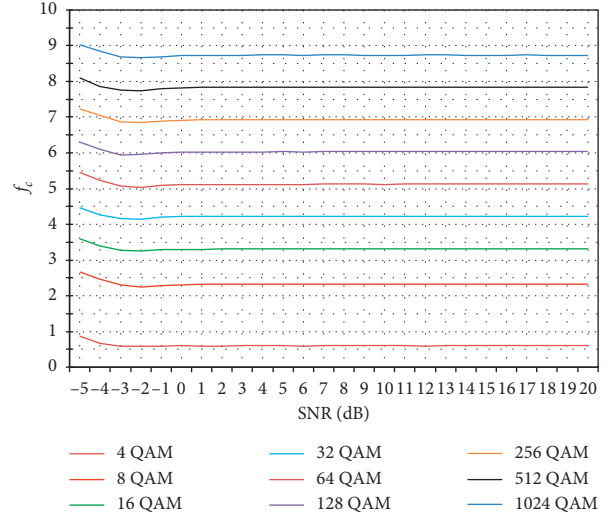

Figure 3: Feature "$f_c$" distribution curve vs SNR with the test of 10,000 signal realizations from {4~1024}-QAM each consisting signal length $N = 4096$ with a phase offset of "$\pi/6$."

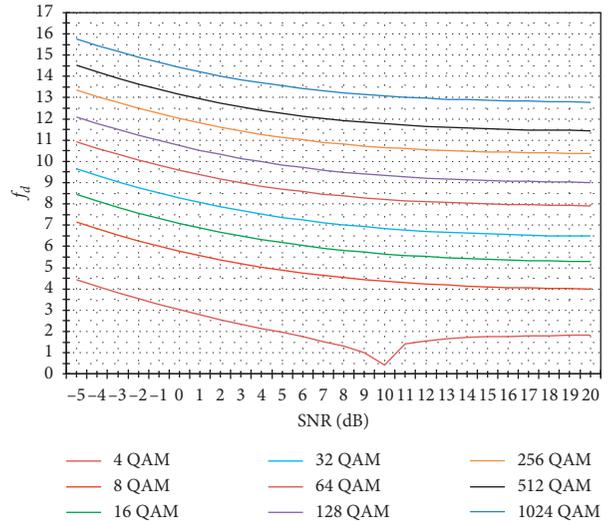

Figure 4: Feature "$f_d$" distribution curve vs SNR with the test of 10,000 signal realizations from {4~1024}-QAM each consisting signal length $N = 4096$ with a phase offset of "$\pi/6$."

Else if $\{f_{x_i} \geq f_x(\varphi_i) \,\&\&\, f_{x_i} \leq f_x(\varphi_i)\}$ $\text{index}_i = i'$, where $2^{n'}$ QAM, $n' > 2$, $i > 1, i' > 1$.

The same procedure is carried out for $\{f_{a_i}, f_{b_i}, f_{c_i}, f_{d_i}\}$ with the increase in the index $i$, until $i = 8$.

Algorithm 2 formally describes the procedure of deciding which classifier has a better probability to discriminate the required modulation format. The outcome of the classifier to recognize a single class is boosted by one, two, or all classifiers. This operation depends on feature distribution curves for each class. Let us suppose that their feature distribution curves have no overlapping parts and the curves are approximately parallel to the SNR axis (within a certain range of SNR). All classifiers set "true." The input is $\{f_{a_i}, f_{b_i}, f_{c_i}, f_{d_i}\}$, with $i$ index.

If $(\text{index} f_{a_i} == i')||(\text{index} f_{b_i} == i')||(\text{index} f_{c_i} == i')||$ $(\text{index} f_{c_i} == i')$ is true, then increase the index $CC_{\text{index SNR}}$. This procedure continues until $\{f_{a_i}, f_{b_i}, f_{c_i}, f_{d_i}\}_{i=8}$.

Algorithm 2 has been developed to shorten the execution time of signal recognition and classifier speed increment and, moreover, to improve the accuracy of the classifier. Algorithm 2 can be summarized as follows: the comparison is achieved by logical operation of the sequential status $\{f_{a_i}, f_{b_i}, f_{c_i}, f_{d_i}\}_i$. As result if either present statue could not recognize the candidate class, in both cases of signal/noise power ratio is greater or less than 0 dB. To overcome that situation, other classifiers strengthened the present status

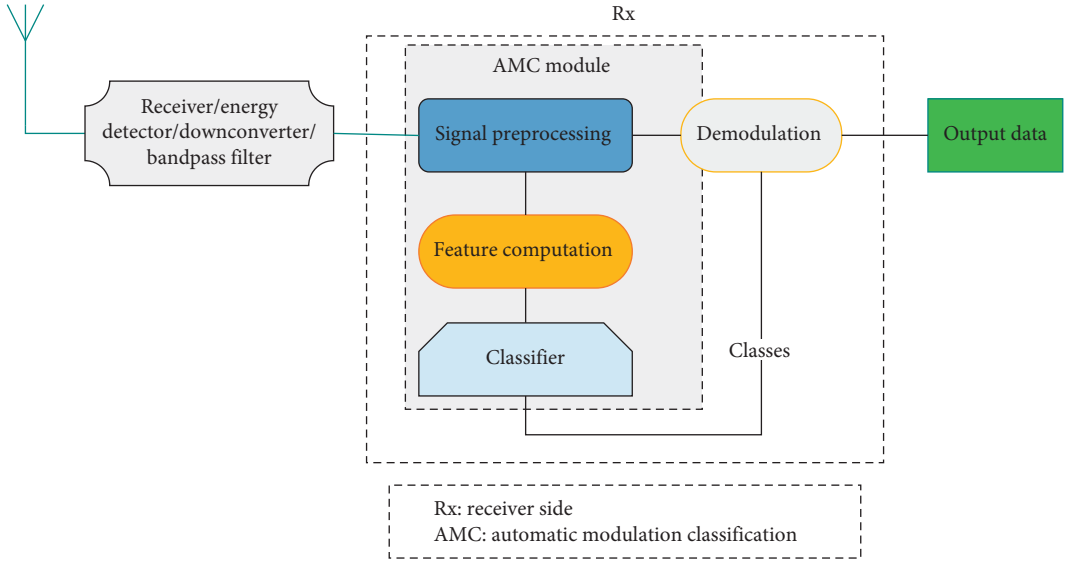

FIGURE 5: Procedure for automatic modulation classification.

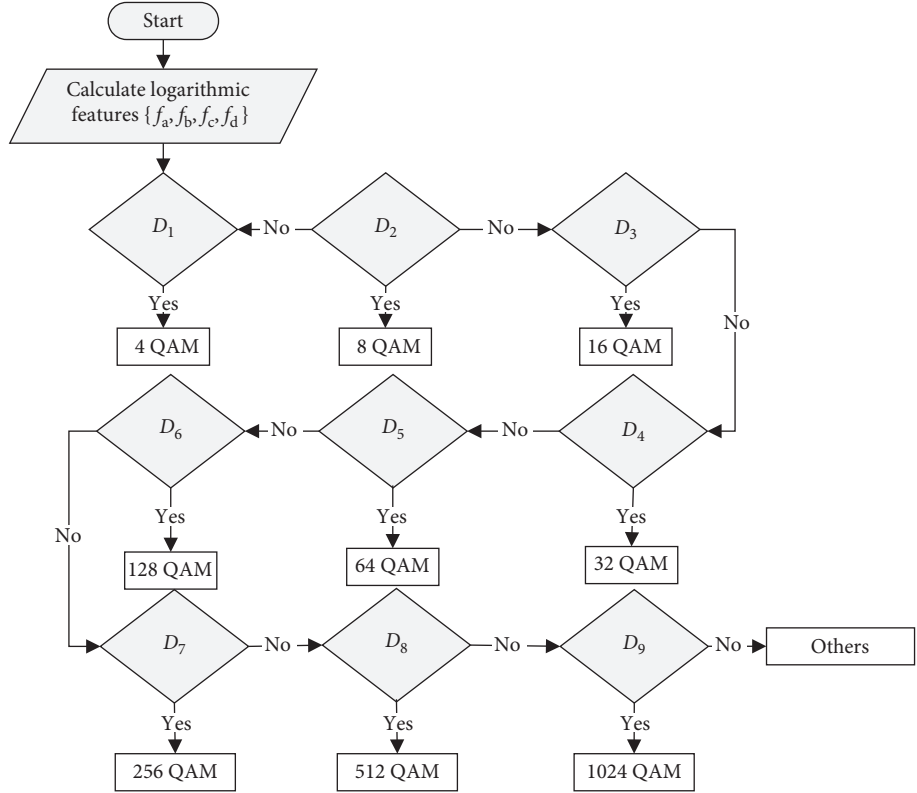

FIGURE 6: Flowchart of the M-QAM recognition algorithm.

and boosted the probability of the correct recognition rate before deciding the correct modulation.

The mathematical expression of the decision boundary of the logarithmic classifier algorithm of the higher order QAM signals is expressed below. For instance, for an input feature $D_{(1)}^x$, the threshold levels are concluded from the results in Table 1 at $i = 0$ and expressed as logical status:

$$D_{(1)}^x = [f_a \geq f_a(\varphi_0) \& f_a < f_a(\varphi_2)]. \tag{11}$$

*Remark:* $i' = (i + 1)$ for 4-QAM, $i' = 1$; likewise the expression of $D_{(i)>1}^x$ is as follows:

$$D_{(i+1)}^a = [f_a \geq f_a(\varphi_{i-1}) \& f_a < f_a(\varphi_{i+1})], \tag{12}$$

$$D_{(i+1)}^b = [f_b \geq f_b(\varphi_{i-1}) \& f_b < f_b(\varphi_{i+1})], \tag{13}$$

$$D_{(i+1)}^c = [f_c \geq f_c(\varphi_{i-1}) \& f_c < f_c(\varphi_{i+1})], \tag{14}$$

TABLE 1: Optimum threshold values to each logarithmic feature.

| I | $f_a$ | $f_b$ | $f_c$ | $f_d$ |
|---|-------|-------|-------|-------|
| 0 | 0.5 | 1.5 | 1.5 | 3 |
| 1 | 0.95 | 2.2 | 3 | 5 |
| 2 | 1.3 | 2.7 | 4 | 6.4 |
| 3 | 1.52 | 3.1 | 5 | 7.6 |
| 4 | 1.8 | 3.7 | 6.3 | 9 |
| 5 | 2.2 | 4.3 | 6.6 | 10 |
| 6 | 2.5 | 4.8 | 7.5 | 11 |
| 7 | 3 | 5.7 | 9 | 12 |
| 8 | 4.5 | 7.5 | 12.5 | 15.5 |

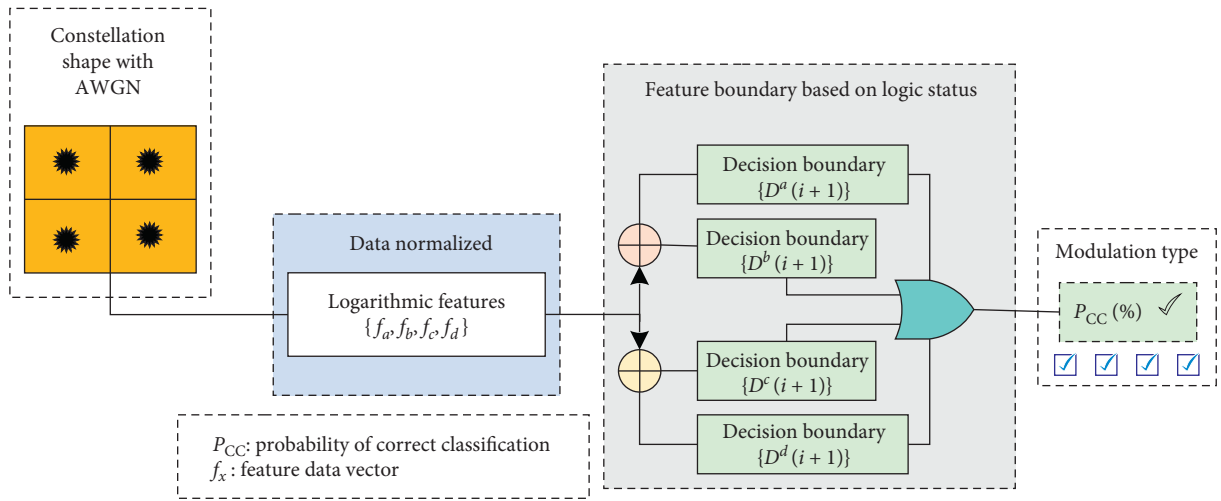

FIGURE 7: The block diagram of our logarithmic classification system.

Input: generate digital M-QAM modulation signals;
$Dx = d_1, d_2, d_3, \ldots, d_i$, and $x \in \{a, b, c, d\}$,
$M = \{2^n\}$, $n = 2, 3, 4, 5, 6, 7, 8, 9, 10$
$s_i = d_x$; $s_i$ row of data after modulated by M-QAM signal
Output: $v \in$ index $\{f_{a_i}, f_{b_i}, f_{c_i}, f_{d_i}\}$; output boundary with index$_i$
Step 1: $s_i = d'_x$; corrupted by Gaussian random noise
Calculate cumulants $C_{11}$, $C_{22}$, $C_{33}$, and $C_{44}$
Step 2: $f_{a_i} = \log_{10}[|C_{11_i}(s_i)|]$; $f_{b_i} = \log_{10}[|C_{22_i}(s_i)|]$;
$f_{c_i} = \log_{10}[|C_{33_i}(s_i)|]$; $f_{d_i} = \log_{10}[|C_{44_i}(s_i)|]$ ;
Step 3: If $\{f_{a_0} < f_x(\varphi_0)\}$; 4-QAM
index$_i$ = 1;
Step 4: Else If $\{f_{x_j} \geq f_x(\varphi_i) \, \&\& \, f_{x_i} \leq f_x(\varphi_i)\}$;
where $2^{n^i}$ QAM; $n^i > 2$, $i > 1$, $i' > 1$
index$_i$ = i';
Step 5: Return $v$;

ALGORITHM 1: Logarithmic classifier of higher order QAM modulation signals.

Input: $v \in$ index $\{f_{a_i}, f_{b_i}, f_{c_i}, f_{d_i}\}_i$; input boundary with index$_i$
Output: $P_{CC}$; probability of correct recognition
Step 1: If (index$f_{a_i}$ == i')$\{||$(index$f_{b_i}$ == i')$||$(index$f_{c_i}$ == i')$||$(index$f_{d_i}$ == i')$\}$;
$CC_{\text{index SNR}} = CC_i + 1$; correct rate counts
End
Step 2: Return $P_{cc}$;

ALGORITHM 2: Choose a higher and better classifier.

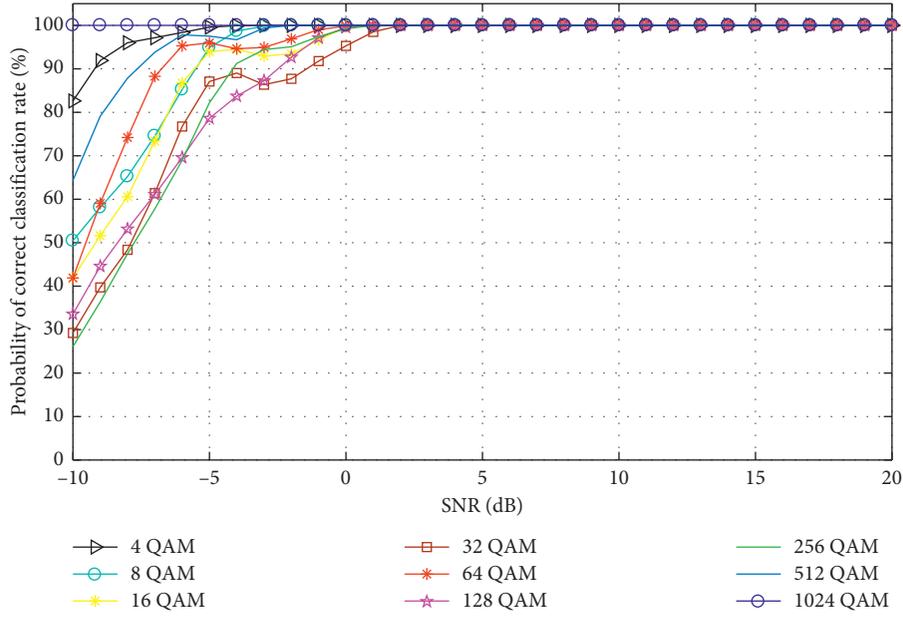

FIGURE 8: The probability of correct classification rate vs SNR with 10,000 signal realizations from {4~1024}-QAM each consisting signal length $N = 4096$ and phase offset "$\pi/6$."

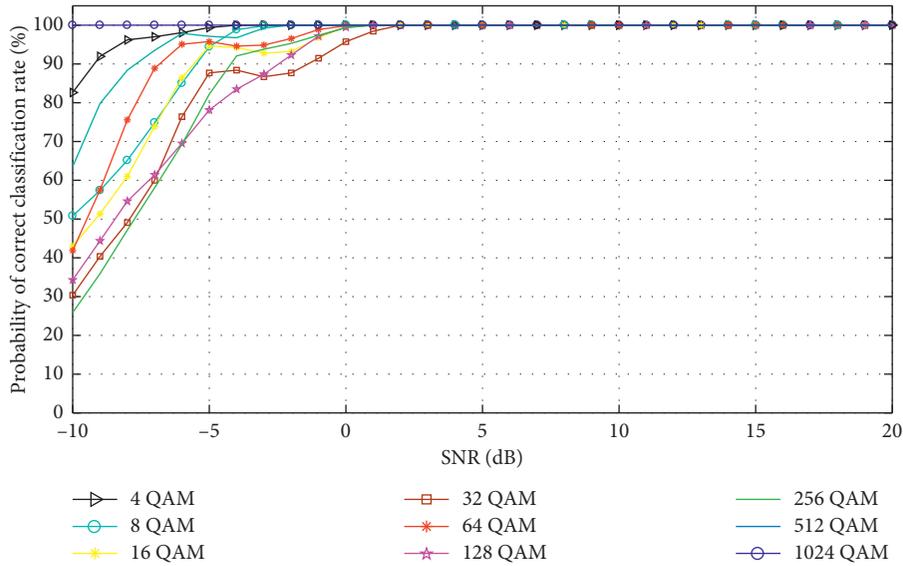

FIGURE 9: The probability of correct classification rate vs SNR with 10,000 signal realizations from {4~1024}-QAM each consisting signal length $N = 4096$ and phase offset "$\pi/4$."

$$D^d_{(i+1)} = [f_d \geq f_d(\varphi_{i-1}) \,\&\, f_d < f_d(\varphi_{i+1})], \quad (15)$$

$$D_i = D^a_{(i+1)} \left\| D^b_{(i+1)} \right\| \left\| D^c_{(i+1)} \right\| \left\| D^d_{(i+1)}, \quad (16)$$

where $f_x(\varphi_i)$ is the threshold value to each of the logarithmic features, in which $x \in \{a, b, c, d\}$ and $i = 0, 2, 3, 4, \ldots, 8$, where the threshold count here is 8, and also $i' = 1, 2, 3, 4, \ldots, 9$ is a class pattern number of QAM signals that is being recognized.

Referring to the flowchart in Figure 7, and equations (12)–(16), under the effect of SNR, the Monte Carlo simulation was carried 10000 times realization for each one of the nine

M-QAM signals and then it was tested by using the proposed logarithmic classifier, and the phase was offset by using three degrees starting from $1/6\pi$ as the reference point.

This work emerges some promising results, and the simulation results are arranged in Figures 8–15. The most striking result that formed in Figures 8–11 shows that when SNR $\geq 5$, the recognition rate of M-QAM signals can reach over 99% even with variation in both sample length and phase offset; therefore, the algorithm proposed in this paper is very efficient in recognizing these signals, and in order to evaluate the performance of the proposed algorithm, a comparison is conducted in terms of the probability of

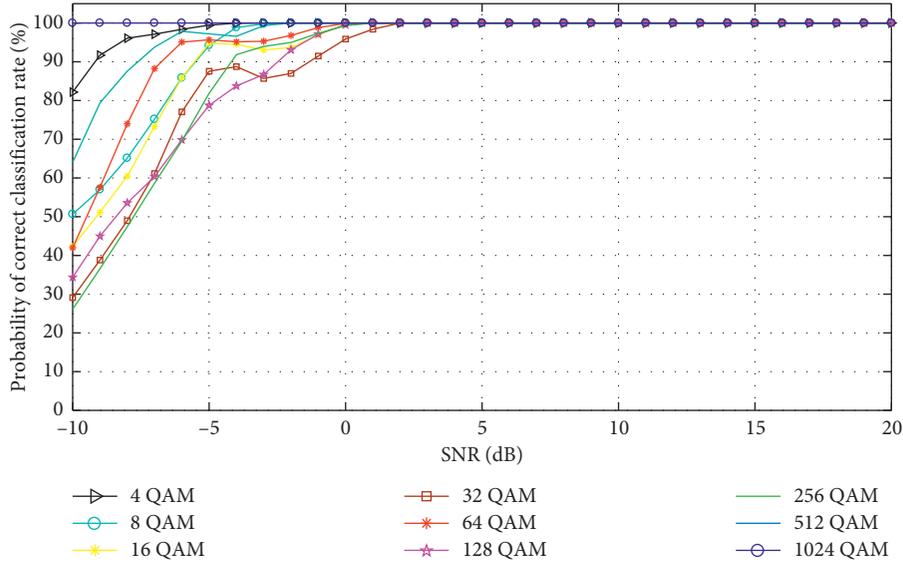

FIGURE 10: The probability of correct classification rate vs SNR with 10,000 signal realizations from {4∼1024}-QAM consisting signal length $N = 4096$ and phase offset "$\pi/3$".

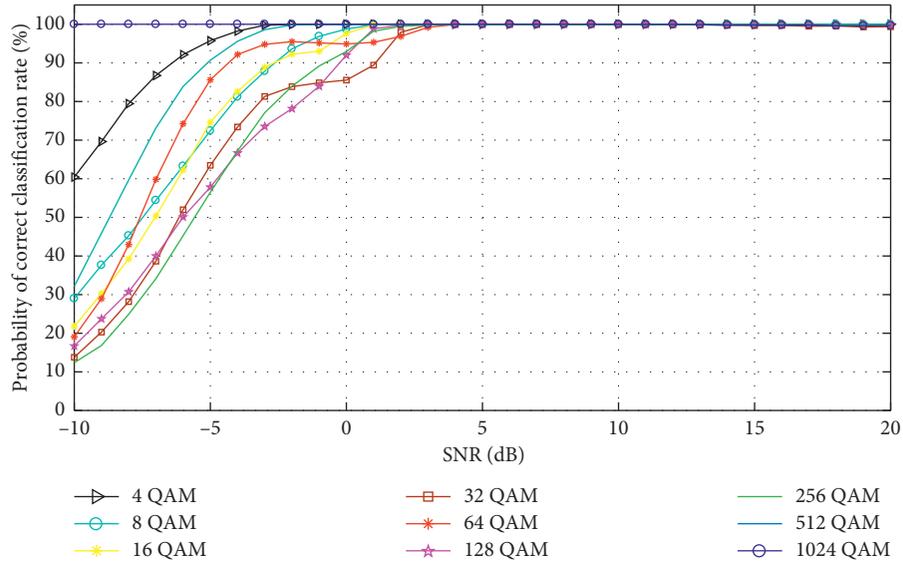

FIGURE 11: The probability of correct classification rate vs SNR with 10,000 signal realizations from {4∼1024}-QAM each consisting variable signal length from N = {64 128 265 512 1024 2048 4096} and phase offset "$\pi/6$".

correct classification (PCC) versus SNR which can be formulated as

$$P_{CC} = \sum_{k=1}^{j} p(\omega_k \mid \omega_k) P(\omega_k),$$

probability of correct classification rate %

$$= P_{CC} \cdot \%100.$$

(17)

In the same direction, "J" is the modulation signal candidate classification $(\omega_1, \omega_2, \omega_3, \omega_j)$, where $P(\omega_k)$ is the probability of the modulation scheme when $(\omega_k)$ occur, while $P(\omega_k \mid \omega_k)$ is the probability of correct classification when an $(\omega_k)$ constellation is sent [32].

## 5. Discussion

All the simulations have been done under the computer simulation environment. The experiments were carried out in three stages in the presence of AWGN: first, extract the HOC features and then pass through the feature modifier, the latter will be a logarithmic classifier based on HOC. Second, make a decision based on the threshold algorithm, and third, achieve the evaluation of PCC % for each modulation type.

It appears from the simulation results that the best performance curve is achieved in using the length of sample; 4096 samples. This was performed based on the fact that increasing the data set can overcome the overfitting problem. Nevertheless, this performance has the highest

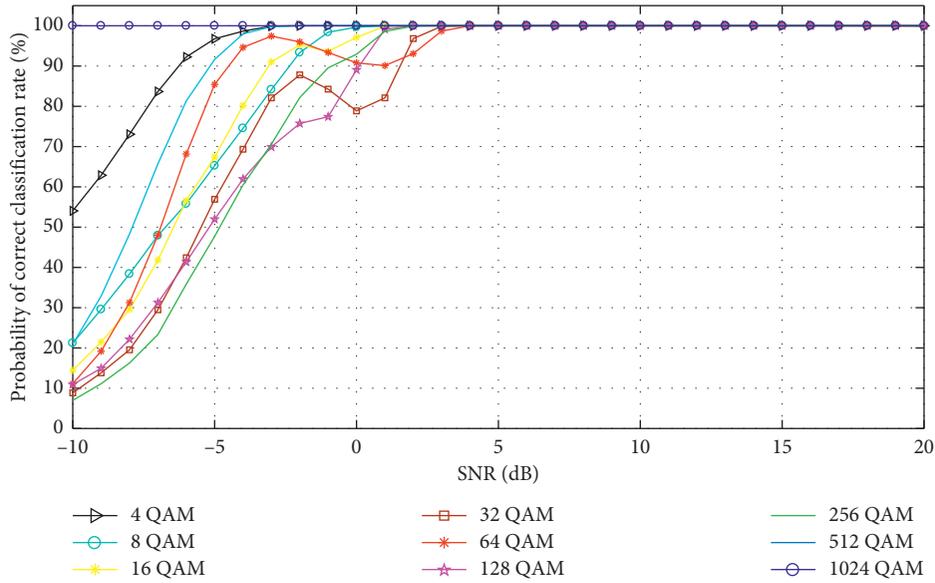

Figure 12: The probability of correct classification rate vs SNR with 10,000 signal realizations {4~1024}-QAM each consisting signal length $N = 256$ and phase offset "$\pi/6$".

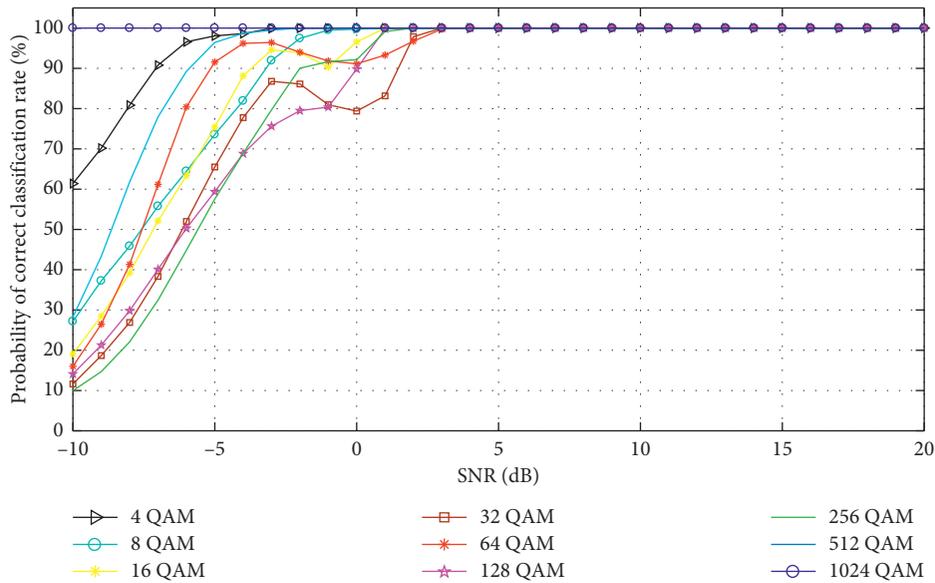

Figure 13: The probability of correct classification rate vs SNR with 10,000 signal realizations from {4~1024}-QAM each consisting signal length $N = 512$ and phase offset "$\pi/6$."

classification rate at a low SNR from −3 dB to 20 dB, as shown in Figure 8.

On the contrary, the curves for variable sample length [64, 128, 256, 512, 1024, 2048, and 4096] indicate a converged accuracy in the same test range of the SNR, and this can be clearly observed in Figure 11.

The accuracy of classifications is lower when reducing the number of samples transmitted, while in the proposed algorithm, the sample length has been reduced to [2048, 1024, 512, 256]. The probability of the correct classification ratio ranges from about 80% to 100% in the SNR range of 0 to 20 dB, as shown in Figures 12–15, respectively. This is mainly due to the reason that we proposed a method which considered that the HOC's feature is based on logarithmic modification; this will create a new classifier known as a logarithmic classifier. However when SNR ≥ −5, the classifier accuracy deteriorates but does not collapse the correct classification rate that becomes below 75% and the phase offset will be considered as shown in Figure 9 in which the phase offset were $\pi/4$, while in Figure 10, the phase offset is $\pi/3$. The proposed algorithm gives a consistent classification accuracy of about 99.7%, in the range of SNR from 2 dB to 20 dB.

It was observed when the phase offset changed; the probability of the correct classification rate is almost equal, and this was due to the high robustness of the proposed classifier. Eventually, the superior performance of the proposed classifier changed the phase offset, and also variability of the transmitted signal length is attributed to the efficient of the proposed logarithmic classifier.

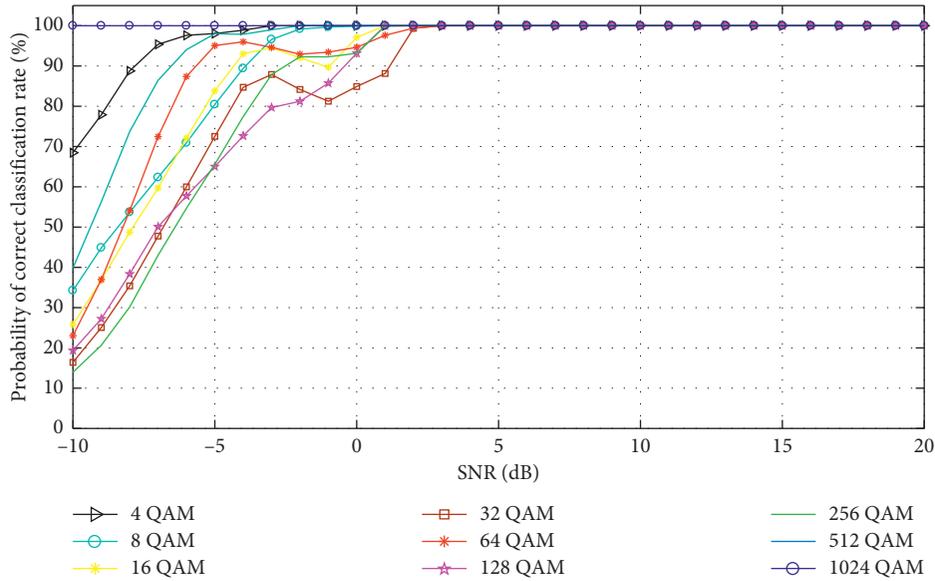

Figure 14: The probability of correct classification rate vs SNR with 10,000 signal realizations from {4-1024}-QAM each consisting signal length $N = 1024$ and phase offset "$\pi/6$."

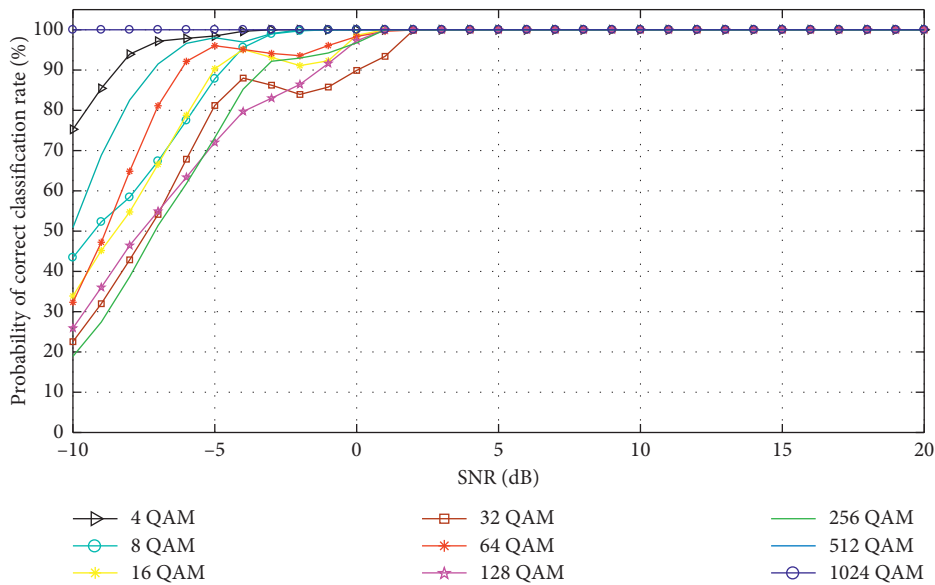

Figure 15: The Probability of correct classification rate vs SNR with 10,000 signal realizations {4-1024}-QAM each consisting signal length $N = 2048$ and phase offset "$\pi/6$."

### 5.1. Evaluation Accuracy and Complexity with Other Methods.
Because of the complexity of the classification algorithm, this research suggests a different method of classifications to conventional modulation [10, 11], which requires a different logarithmic operations. Yet, still it is crucial in terms of high probabilities to correct the recognition rate and less process is required to calculate the cumulant orders; at the same time, the other studies do not highlight these promising lines [21, 23, 33–35].

Although cumulant-based classifier methods were convenient to cope with M-QAM modulation recognition, yet it does not provide a sufficient classification rate M-QAM modes under fading channel, and it can even produce missing classification in some circumstances, while the logarithmic classifier of M-QAM signals has a better probability of correct recognition rate at SNR range between −5 dB and 20 dB.

The performance of the logarithmic classifier of M-QAM is superior in what was proposed in terms of its ability to distinguish between high order and very high order of QAM signals and thus can be considered as one of the pillars for the modulation techniques in the future due to unprecedented demand for the wireless communication

Table 2: Comparison between proposed classifier with other systems in the literature.

| Reference | Classifier structure | | Modulation type | | System performance | | Simulation tool |
| | Features | Complexity | M-QAM | M-PSK | SNR (dB) | Accuracy (%) | |
|---|---|---|---|---|---|---|---|
| [33] | $C_{40}, C_{41}, C_{42}, C_{60},$ $C_{61}, C_{62}, C_{63},$ | High | 16, 64 | 4 | −3 | Not covering | |
| | | | | | 0 | Not covering | |
| | | | | | ≥+5 | 78.4–100 | |
| [34] | $C_{40}, C_{41}, C_{42}, C_{60},$ $C_{61}, C_{62}, C_{63},$ | High | 16, 64 | 2, 4 | −3 | Not covering | |
| | | | | | 0 | Not covering | |
| | | | | | ≥+5 | 89.8–100 | |
| [11] | $C_{11}, C_{22}, C_{33}, C_{44},$ | Medium | 16~256 | No | −3 | 98.33 | |
| | | | | | 0 | 100 | |
| | | | | | ≥+5 | 100 | |
| [35] | $C_{20}, C_{21}, C_{40}, C_{41},$ $C_{42}, C_{60}, C_{61}, C_{62},$ $C_{63}$ | High | 16, 64, 256 | 2, 4, 8 | −3 | Not covering | |
| | | | | | 0 | 77.6 | MATLAB functions are invoked to implement and evaluate the performance |
| | | | | | ≥+5 | 99.96 | |
| [23] | $C_{20}, C_{21}, C_{40}, C_{41},$ $C_{42}, C_{60}, C_{61}, C_{62},$ $C_{63}, C_{80}, C_{81}, C_{82},$ $C_{83}, C_{84}$ | High | 16, 64, 256 | 2, 4, 8, and others | −3 | Not covering | |
| | | | | | 0 | Not covering | |
| | | | | | ≥+5 | Unknown | |
| [21] | $C_{20}, C_{21}, C_{41}, C_{42},$ $C_{63}$ | Medium | 16, 64, V.29 | 4 | −3 | 60 | |
| | | | | | 0 | 81 | |
| | | | | | ≥+5 | 90 | |
| [10] | $C_{20}, C_{21}, C_{40}, C_{41},$ $C_{42}$ | Medium | 8, 16, 32, 64 | 4, 8, 16, 32, 64 | −3 | 26 | |
| | | | | | 0 | 45 | |
| | | | | | ≥+5 | 95 | |
| Proposed algorithm | Logarithmic cumulant | Medium | (8~1024) | 4 | −3 | 75–95 | |
| | | | | | 0 | 80–98 | |
| | | | | | ≥+5 | 100 | |

"~" refers values from M-QAM to M-QAM and "," refers values M-QAM and M-QAM; note that signal length $N = 4096$.

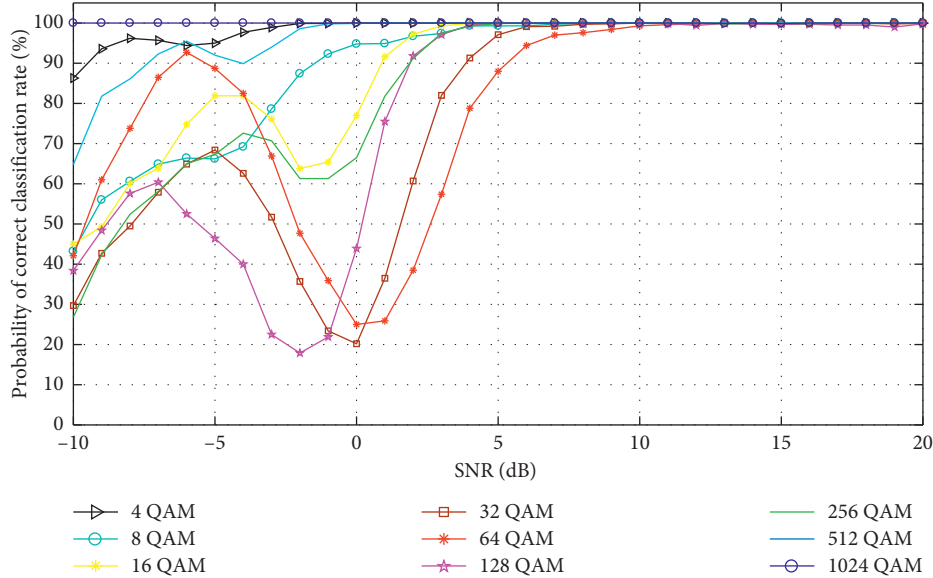

Figure 16: The probability of correct classification rate vs SNR with 10,000 signal realizations from {4~1024}-QAM each consisting signal length $N = 4096$ and phase offset "$\pi/6$" under AWGN plus fast, frequency selective Ricean fading channel with $f_D = 5$ kHz.

technology expecting to reach an estimated 7.6 billion by 2020 [36].

According to Table 2, the fourth-order cumulant classifier has the lowest complexity, while the combination between forth-, sixth-, and eighth-order cumulants based classifier has the highest complications due to the use of exponential operations; however, it is worth to clarify that the logarithmic classifier in this research work depends on unique combination between lowest and highest orders of cumulant, and this combination will create an integrity and a less complexity classifier system. The complexity of several methods proposed that occurred in previous work was compared in detail as shown in Table 2.

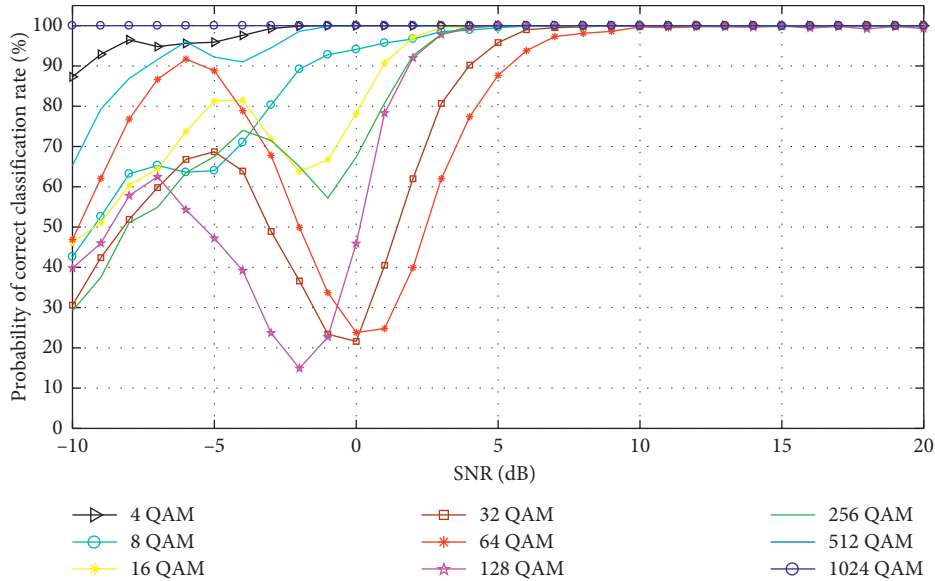

Figure 17: The probability of correct classification rate vs SNR with 10,000 signal realizations from {4~1024}-QAM each consisting signal length $N = 4096$ and phase offset "$\pi/6$" under AWGN plus fast, frequency selective Rayleigh fading channel with $f_D = 5$ kHz.

The results shown in Figures 16 and 17 indicate the performance degradation of classification for QAM signals versus their performance in the AWGN channel. It is not difficult to understand that the cause of the performance degradation comes from the multipath channel fading and Doppler fading effects model. The weakness performance of cumulant-based classifiers with fading channels has been documented in [22]. The only exception in the logarithmic classifier that maintains the same level of classification accuracy is typically at SNR >5 dB despite the different fading channel models used.

## 6. Robustness Evaluation Cases

### 6.1. Under Different Channel Models.
Research in the modulation classification of digital signals approximately started since forty years ago, and the robustness evaluation of conventional AMC methods under the AWGN channel has been well established in the literature [37–39]. Similarly, the use of AWGN as the channel model is a common practice for communication engineers and researchers [19, 40].

In information theory, the basic communication channel model AWGN that represents the effect of several random processes occurring in nature is in the same direction, by which AWGN is used as a channel model where the communication impairment is an addition of white noise with a constant spectral density and normally distributed amplitude. In this work, the proposed logarithmic classifier has shown superior performance under the AWGN channel even with low SNR, also with variable lengths of the transmitted signal.

Table 2 shows the robustness analysis among previous works which were investigated in the development of the modulation classification algorithm. It appears very clearly that the logarithmic classifier has higher classification accuracy of modulation schemes {4-1024}-QAM, which also achieves a high percentage when the SNR ≥ 5.

Although AWGN alone could be a proper model to the degradation of signals in free space, yet the existence of physical obstructions such as building and towers can cause multipath propagation losses. It implies that the signal reaches the receiver site via multiple propagation paths. Respecting to the relative lengths of the paths, this could be either constructive or destructive interference. Furthermore, relative motion between the transmitter and receiver may cause the Doppler shift. The performance of the proposed classifier is also investigated in the AWGN channel with fast frequency-selective fading also with the effect of Rayleigh and Rician magnitude channel fading, while the phase was randomly distributed and was assumed to be constant through the symbol period, which indicates no loss in signals. Rician factor is defined as the ratio between the power of the direct path and the power of the reflected paths, which is a constant value. Here, it is assumed to be 3 dB.

The specification used was based on Molisch and his coworker research [41], with a symbol rate of $3.84 \times 10^6$ symbols per second, average path gains [0, −0.9, −4.9, −8, −7.8, −23.9]dB, path delays $[0, 2, 8, 12, 23, 37] \times 10^{-7}$ seconds, and maximum Doppler shift of 5 kHz. Figures 16 and 17 depict the probability of correct recognition rate at signal length $N = 4096$ in the AWGN plus fast, frequency selective with 50 kHz of maximum Doppler shifts with Rician and Rayleigh fading channels, respectively. Furthermore, Figures 18 and 19 show the probability of correct recognition rate with variable signal lengths in AWGN plus fast, frequency selective with 50 kHz of maximum Doppler shifts in Rician and Rayleigh fading channels, respectively.

As could be observed in Figures 16 and 17, respectively, under signal propagation over multipath, it gives an accepted classification rate when SNR is not lower than 4 dB, although with presence of effect of the Doppler shift. Also, it could be observed in Figures 16 and 17. The probability of the correct classification rate at SNR > 4 dB is higher for the Rician fading channels compared to the Rayleigh fading channels. This is because of the existence of the strong signal due to the line of sight component. It is well noted that a higher Doppler shift results in a lower probability of correct classification [41]. As well as contrast to that, the logarithmic classifier provides unexpected performance even with the existence of the Doppler shift, and signal is transmitted with a variable length between {64-4096}. It is also important to underline that the proposed classifier vouchsafes satisfactory results under both fading channel environments Rician and Rayleigh.

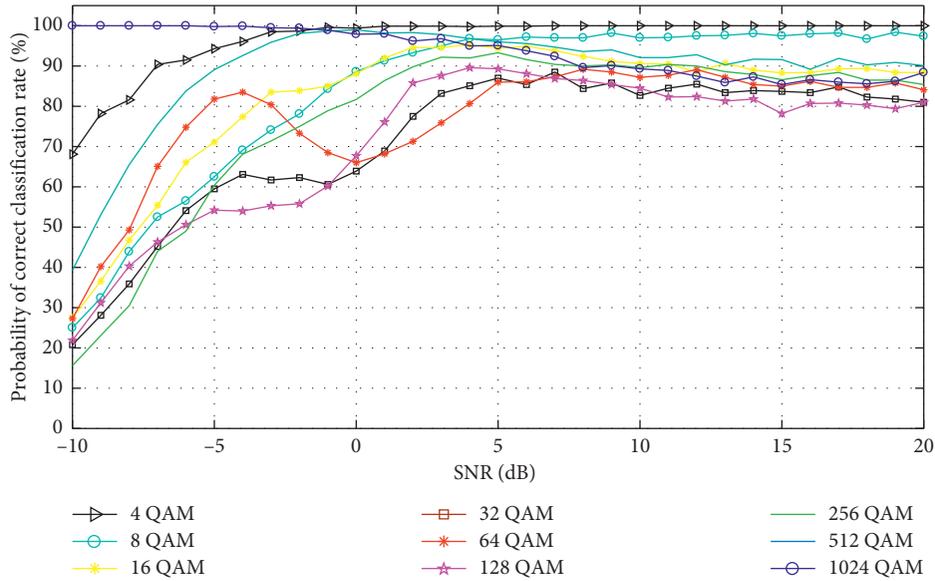

FIGURE 18: The probability of correct classification rate vs SNR with 10,000 signal realizations from {4~1024}-QAM each consisting variable signal length from $N$ = {64 128 265 512 1024 2048 4096} and phase offset "$\pi/6$" under AWGN plus fast, frequency selective Ricean fading channel with $f_D$ = 5 kHz.

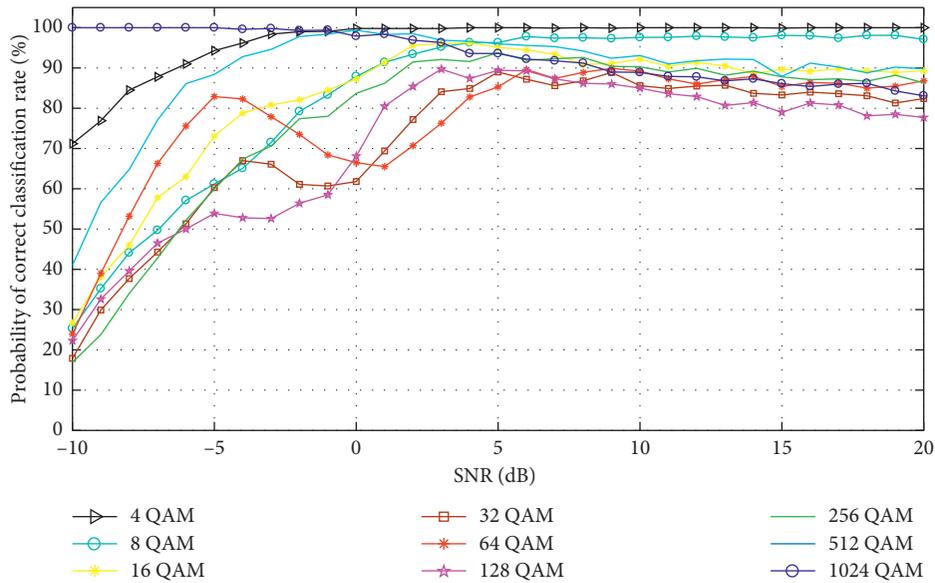

FIGURE 19: The probability of correct classification rate vs SNR with 10,000 signal realizations from {4~1024}-QAM each consisting variable signal length from $N$ = {64 128 265 512 1024 2048 4096} and phase offset "$\pi/6$" under AWGN plus fast, frequency selective Rayleigh fading channel with $f_D$ = 5 kHz.

As could also be observed in Figures 16 and 17, mainly, at SNR lower than −2 dB, the correct classification rate has fallen below 80%, especially to {32, 64, 128, 256} QAM signals. This could be noted when the SNR is lower than 0 dB.

The intraclass recognition of the modulation order using the logarithmic classifier gives different results depending on the modulation type. For example, the simulations in this work illustrated that this recognition will be better for 1024-

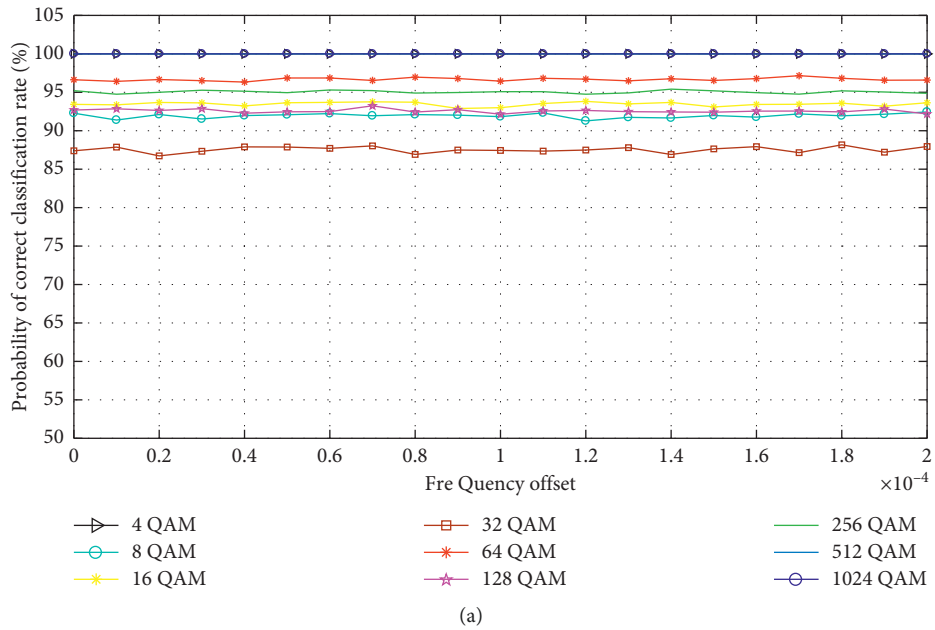

(a)

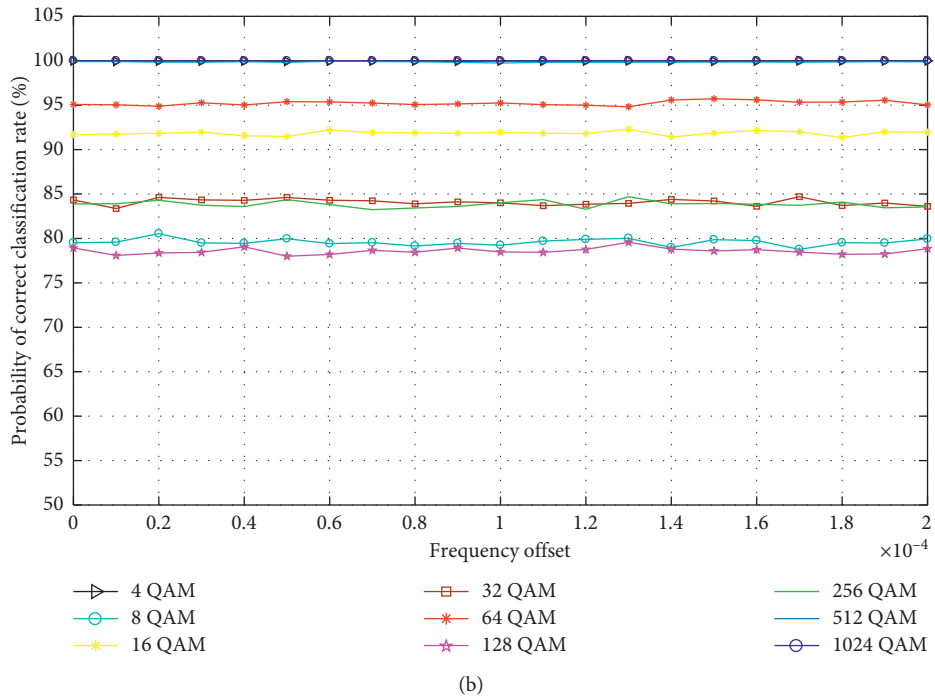

(b)

FIGURE 20: (a) Probability of the correct classification rate vs frequency offset and signal length $N = 4096$. (b) Probability of the correct classification rate vs frequency offset with variable signal lengths from $N = \{64\ 128\ 265\ 512\ 1024\ 2048\ 4096\}$ with a phase offset of "$\pi/6$." $\{4\sim1024\}$-QAM with 10,000 signal under AWGN—only SNR $= -2$ dB.

QAM and 512-QAM schemes than other modulation schemes, where the probability percent is high to correct the classification rate. Similarly, the SNR is not lower than 0 dB, and the results show that the probability of the recognition rate gives the lowest percentage for 32-QAM and 64-QAM signals, which can be observed in Figures 16–19. It is also important to mention that the fluctuations in performance in the correct recognition rate is in higher order QAM signals, and that is due to the internal structures of the logarithmic classifier.

It seems that the effect of variation in the frequency offset of the transmitted signals almost insignificantly influences the logarithmic classifier although under SNR $= -2$ and random signal lengths. Figures 20(a) and 20(b) depict the probability of the correct classification rate vs frequency offset and signal length $N = 4096$. Likewise Figure 20(b) depicts the probability of the correct classification rate vs frequency offset with a variable signal length and the signals corrupted by AWGN. This end could be an important point to the proposed classifier among the AMC methods that have tested the influence of the offset frequency to the transmitted signal. In the same direction, Figure 21 shows a change in the number of transmitted symbols vs probability of the correct classification rate. It also well-observed that the proposed method has a better classification rate than the proposed algorithm does [23]. Even with this, it is a very difficult circumstance with SNR = −2 dB in the AWGN channel. Eventually, the logarithmic classification algorithm is not the best one. On the contrary, it facilitated the process of classification of higher order QAM signals. Nevertheless, degradation points of the logarithmic classifier are that it is a channel-dependent model. Any variations in the type of noise (colored vs. AWGN), the number of transmitted signals present in the channel, and the type of possible modulated signals require an overhaul of all four tree threshold levels.

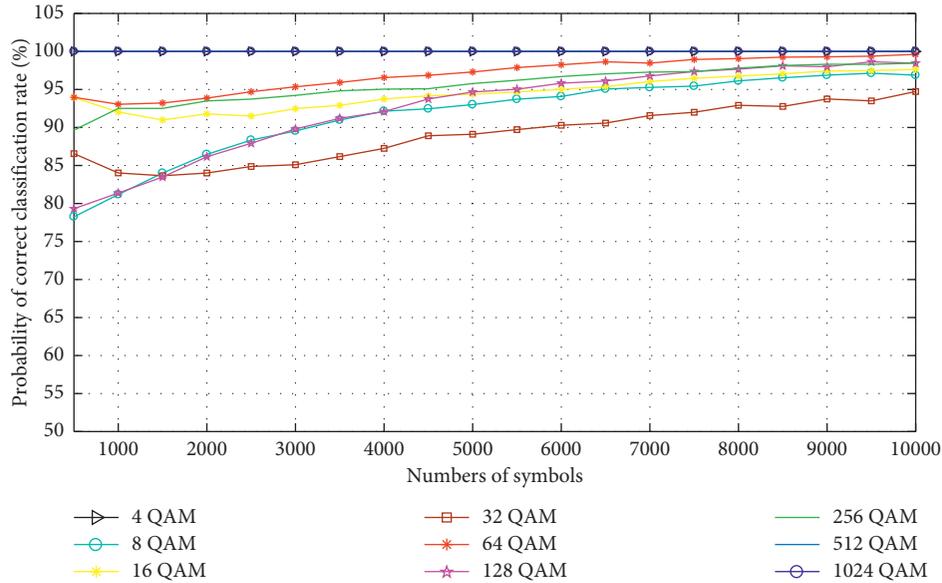

Figure 21: The probability of the correct classification rate vs the number of symbols SNR with 10,000 signal realizations from {4~1024}-QAM, phase offset "$\pi/6$", and signal degradation using AWGN—only SNR = −2 dB.

## 7. General Trends

Unlike the traditional AMC algorithms, the logarithmic classifier has functional features, which make them the preferable classifier for the medium and higher order of M-QAM modulation schemes. Likewise, the logarithmic classifier is very useful for channels that only receive corruption by AWGN. The prominent results related to the AWGN channel without fading effect have shown unexpected performance. Although the classification of higher order modulation signals is more challenging than to lower order signals due to the distance between the constellation points is small, but on the contrary, it appears nonreal convergence during the evaluation of the robustness of the logarithmic classifier. The probability of correct classification suffered from disparity especially for lower order modulation schemes at the expense of schemes that have smaller constellation point distance; therefore from this end, it might be concluded that more research is needed on the logarithmic classifier particularly SNR = 0. Although cumulant-based classifier schemes with many variations is the convenient way to cope with M-QAM modulation schemes, signal cumulant features, or uncoordinated combination between cumulants and moments does not provide sufficient classification under multipath fading channels and can even produce contrast in the classification rate. It is also notable that this classifier technique is not unique in terms of fixed threshold and there exist various ways to determine threshold levels automatically in order to improve the classification sense to the desired modulation schemes. However, the comprehensive review reveals that the logarithmic classifier with fixed threshold has overcome all weaknesses points even with a severe noisy environment, moreover, under multipath propagation channels. On the contrary, this work has provided references that help researchers make progress toward in the direction of developed new generation of AMC systems, which is called as the logarithmic classifier.

## 8. Conclusion

This paper provides a method that transacts with the modulation recognition of higher order M-QAM signals under the effect of AWGN on a channel environment through the usage of classifier which is cumulant-based natural logarithmic characteristics with fixed thresholds.

The combination of higher order cumulant and properties of the logarithmic functions created a new generation of the cumulant-based classifier, who provides a superior performance to classify the higher order M-QAM signals even with a low range of SNR.

The simulation results indicate that the correct classification rate can reach over 92% at an SNR of 4 dB under the AWGN, even as it seems moderate over different fading channels.

According to a comparison with the recently proposed algorithms in the literature, the performance of the logarithmic classifier was efficient and also less complex.

## Abbreviations

| | |
|---|---|
| AMC: | Automatic modulation classification |
| HOM: | Higher order moments |
| HOC: | Higher order cumulant |
| M-QAM: | Multilevel quadrature amplitude modulation |
| AWGN: | Additive white Gaussian noise |
| $N$: | Signal length |
| $C_{xpq}$: | $n^{th}$ order statistical cumulant with q-conjugate order |
| $M_{xpq}$: | $n^{th}$ order statistical moment with q-conjugate order |
| $f_x$: | Logarithmic feature transformer and $x \in \{a, b, c, d\}$ |
| $i$: | Threshold count integer number |
| $i'$: | Number of class pattern status of the modulation scheme |
| $P_{CC}$: | Probability of correct classification |
| $f_D$: | Maximum Doppler shift |
| M~M: | M-QAM ... M-QAM |
| SNR: | Signal-to-noise ratio. |

## Data Availability

No data were used to support this study; thereby, this paper provides a new entry for the M-QAM modulation recognition algorithm. The higher order {4∼1024}-QAM signals are considered, and new attributes were extracted from signals. "MATLAB Communications Toolbox" was used to evaluate the proposed method. Raw data are randomly generated by fetching the random function in the simulation program. For more details, contact the authors.

## Conflicts of Interest

The authors declare that they have no conflicts of interest.

## References

[1] O. A. Dobre, A. Abdi, Y. Bar-Ness, and W. Su, "Survey of automatic modulation classification techniques: classical approaches and new trends," *IET Communications*, vol. 1, no. 2, pp. 137–156, 2007.

[2] Z. Xing and Y. Gao, "Method to reduce the signal-to-noise ratio required for modulation recognition based on logarithmic properties," *IET Communications*, vol. 12, no. 11, pp. 1360–1366, 2018.

[3] J. A. Sills, "Maximum-likelihood modulation classification for PSK/QAM," in *Proceedings of the IEEE Military Communications Conference Proceedings MILCOM 1999*, vol. 1, pp. 217–220, Atlantic City, NJ, USA, October-November 1999.

[4] F. Hameed, O. Dobre, and D. Popescu, "On the likelihood-based approach to modulation classification," *IEEE Transactions on Wireless Communications*, vol. 8, no. 12, pp. 5884–5892, 2009.

[5] W. Wei and J. M. Mendel, "A new maximum-likelihood method for modulation classification," in *Proceedings of the 1995 Conference Record of the Twenty-Ninth Asilomar Conference on Signals, Systems and Computers*, vol. 2, pp. 1132–1136, Pacific Grove, CA, USA, October 1995.

[6] E. E. Azzouz and A. K. Nandi, "Automatic identification of digital modulation types," *Signal Processing*, vol. 47, no. 1, pp. 55–69, 1995.

[7] M. Abdelbar, W. H. Tranter, L. Fellow, T. Bose, and S. Member, "Cooperative cumulants-based modulation classification in distributed networks," *IEEE Transactions on Cognitive Communications and Networking*, vol. 4, no. 3, pp. 446–461, 2018.

[8] J. L. Xu, W. Su, and M. Zhou, "Likelihood-ratio approaches to automatic modulation classification," *IEEE Transactions on Systems, Man, and Cybernetics, Part C (Applications and Reviews)*, vol. 41, no. 4, pp. 455–469, 2011.

[9] H. Hosseinzadeh, F. Razzazi, and A. Haghbin, "A self training approach to automatic modulation classification based on semi-supervised online passive aggressive algorithm," *Wireless Personal Communications*, vol. 82, no. 3, pp. 1303–1319, 2015.

[10] A. Swami and B. M. Sadler, "Hierarchical digital modulation classification using cumulants," *IEEE Transactions on Communications*, vol. 48, no. 3, pp. 416–429, 2000.

[11] I. A. Hashim, J. W. Abdul Sadah, T. R. Saeed, and J. K. Ali, "Recognition of QAM signals with low SNR using a combined threshold algorithm," *IETE Journal of Research*, vol. 61, no. 1, pp. 65–71, 2015.

[12] M. L. Dukiü and G. B. Markoviü, "Automatic modulation classification using cumulants with repeated classification attempts," in *20th Telecommunications Forum (TELFOR)*, vol. 4, pp. 424–427, 2012.

[13] D. S. Chirov, "Application of the decision trees to recognize the types of digital modulation of radio signals in cognitive systems of HF communication," in *Proceedings of the 2018 Systems of Signal Synchronization, Generating and Processing in Telecommunications*, pp. 1–6, Minsk, Belarus, July 2018.

[14] B. G. Mobasseri, "Constellation shape as a robust signature for digital modulation recognition," in *Proceedings of the IEEE Conference on Military Communications MILCOM 1999*, pp. 442–446, Piscataway, NJ, USA, October-November 1999.

[15] B. G. Mobasseri, "Digital modulation classification using constellation shape," *Signal Processing*, vol. 80, no. 2, pp. 251–277, 2000.

[16] C. Yin, B. Li, Y. Li, and B. Lan, "Modulation classification of MQAM signals based on density spectrum of the constellations," in *Proceedings of the 2010 2nd International Conference on Future Computer and Communication ICFCC 2010*, vol. 3, pp. 57–61, Wuhan, China, May 2010.

[17] A. Fehske, J. Gaeddert, and J. H. Reed, "A new approach to signal classification using spectral correlation and neural networks," in *Proceedings of the First IEEE International Symposium on New Frontiers in Dynamic Spectrum Access Networks DySPAN 2005*, pp. 144–150, Baltimore, MD, USA, November 2005.

[18] W. C. Headley, J. D. Reed, and C. R. C. Da Silva, "Distributed cyclic spectrum feature-based modulation classification," in *Proceedings of the 2008 IEEE Wireless Communications and Networking Conference*, pp. 1200–1204, Las Vegas, NV, USA, March-April 2008.

[19] Y. Yuan, P. Zhao, B. Wang, and B. Wu, "Hybrid maximum likelihood modulation classification for continuous phase modulations," *IEEE Communications Letters*, vol. 20, no. 3, pp. 450–453, 2016.

[20] H. T. Fu, Q. Wan, and R. Shi, "Modulation classification based on cyclic spectral features for co-channel time-frequency overlapped two-signal," in *Proceedings of the 2009 Pacific-Asia Conference on Circuits, Communications and Systems PACCS 2009*, pp. 31–34, Chengdu, China, May 2009.

[21] O. A. Dobre, M. Oner, S. Rajan, and R. Inkol, "Cyclo-stationarity-based robust algorithms for QAM signal identification," *IEEE Communications Letters*, vol. 16, no. 1, pp. 12–15, 2012.

[22] O. A. Dobre, A. Abdi, Y. Bar-Ness, and W. Su, "Cyclo-stationarity-based modulation classification of linear digital modulations in flat fading channels," *Wireless Personal Communications*, vol. 54, no. 4, pp. 699–717, 2010.

[23] O. A. Dobre, Y. Bar-Ness, and W. Su, "Higher-order cyclic cumulants for high order modulation classification," in *Proceedings of the 2003 IEEE conference on Military communications MILCOM 2003*, pp. 112–117, Boston, MA, USA, October 2003.

[24] C. M. Spooner, "On the utility of sixth-order cyclic cumulants for RF signal classification," in *Proceedings of the Conference Record-Asilomar Conference on Signals, Systems and Computers*, vol. 1, pp. 890–897, Pacific Grove, CA, USA, November 2001.

[25] F. Ghofrani, A. Jamshidi, and A. Keshavarz-Haddad, "Internet traffic classification using hidden naive Bayes model," in *Proceedings of the ICEE 2015 23rd Iranian Conference on Electrical Engineering*, vol. 10, pp. 235–240, Tehran, Iran, May 2015.

[26] F. Ghofrani, A. Keshavarz-Haddad, and A. Jamshidi, "Internet traffic classification using multiple classifiers," in *Proceedings of the 2015 7th International Conference on Knowledge and Smart Technology (KST) IKT 2015*, Chonburi, Thailand, January 2015.

[27] S. Norouzi, A. Jamshidi, and A. R. Zolghadrasli, "Adaptive modulation recognition based on the evolutionary algorithms," *Applied Soft Computing*, vol. 43, pp. 312–319, 2016.

[28] H. Agirman-Tosun, Y. Liu, A. M. Haimovich et al., "Modulation classification of MIMO-OFDM signals by independent component analysis and support vector machines," in *Proceedings of the 2011 Conference Record of the Forty Fifth Asilomar Conference on Signals, Systems and Computers (ASILOMAR)*, pp. 1903–1907, Pacific Grove, CA, USA, November 2011.

[29] D. Boutte and B. Santhanam, "A hybrid ICA-SVM approach to continuous phase modulation recognition," *IEEE Signal Processing Letters*, vol. 16, no. 5, pp. 402–405, 2009.

[30] B. Ramkumar, "Automatic modulation classification for cognitive radios using cyclic feature detection," *IEEE Circuits and Systems Magazine*, vol. 9, no. 2, pp. 27–45, 2009.

[31] A. Ali, F. Yangyu, and S. Liu, "Automatic modulation classification of digital modulation signals with stacked autoencoders," *Digital Signal Processing*, vol. 71, pp. 108–116, 2017.

[32] S. Xi and H.-C. Wu, "Robust automatic modulation classification using cumulant features in the presence of fading channels," in *Proceedings of theIEEE Wireless Communications and Networking Conference, 2006. WCNC 2006*, pp. 2094–2099, Las Vegas, NV, USA, April 2006.

[33] Z. Zhu, M. Waqar Aslam, and A. K. Nandi, "Genetic algorithm optimized distribution sampling test for M-QAM modulation classification," *Signal Processing*, vol. 94, no. 1, pp. 264–277, 2014.

[34] N. Ahmadi and R. Berangi, "Modulation classification of QAM and PSK from their constellation using Genetic Algorithm and hierarchical clustering," in *Proceedings of the 2008 3rd International Conference on Information & Communication Technologies: from Theory to Applications*, vol. 11670, Damascus, Syria, April 2008.

[35] A. Abdelmutalab, K. Assaleh, and M. El-Tarhuni, "Automatic modulation classification based on high order cumulants and hierarchical polynomial classifiers," *Physical Communication*, vol. 21, pp. 10–18, 2016.

[36] M. H. Alsharif and R. Nordin, "Evolution towards fifth generation (5G) wireless networks: current trends and challenges in the deployment of millimeter wave, massive MIMO, and small cells," *Telecommunication Systems*, vol. 64, no. 4, pp. 617–637, 2017.

[37] C.-S. Park, J.-H. Choi, S.-P. Nah, W. Jang, and D. Y. Kim, "Automatic modulation recognition of digital signals using wavelet features and SVM," in *Proceedings of the 2008 10th International Conference on Advanced Communication Technology*, vol. 1, pp. 387–390, Gangwon-Do, South Korea, February 2008.

[38] N. Ahmadi, "Using fuzzy clustering and TTSAS algorithm for modulation classification based on constellation diagram," *Engineering Applications of Artificial Intelligence*, vol. 23, no. 3, pp. 357–370, 2010.

[39] L. Zhou, Z. Sun, and W. Wang, "Learning to short-time Fourier transform in spectrum sensing," *Physical Communication*, vol. 25, pp. 420–425, 2017.

[40] M. H. Valipour, M. M. Homayounpour, and M. A. Mehralian, "Automatic digital modulation recognition in presence of noise using SVM and PSO," in *Proceedings of the 6th International Symposium on Telecommunications IST 2012*, pp. 378–382, Tehran, Iran, November 2012.

[41] A. F. Molisch, K. Balakrishnan, D. Cassioli, and C.-C. Chong, "IEEE 802.15.4a channel model-final report," *Environments*, pp. 1–40, 2005.

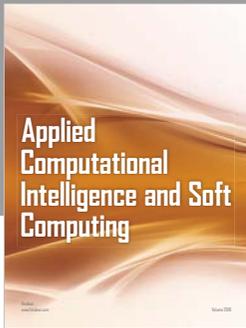

**Applied Computational Intelligence and Soft Computing**

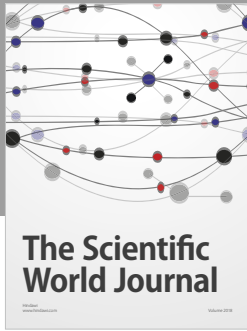

**The Scientific World Journal**

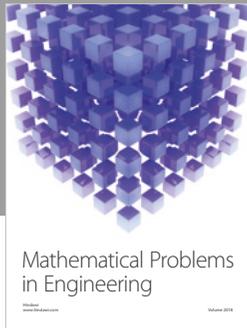

Mathematical Problems in Engineering

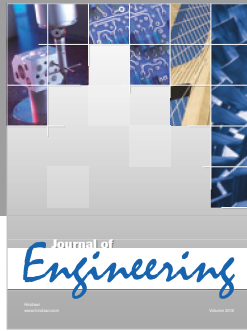

Journal of Engineering

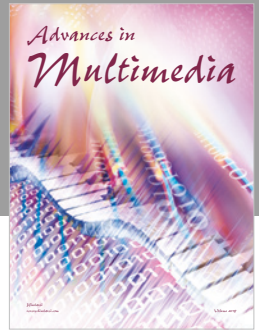

Advances in Multimedia

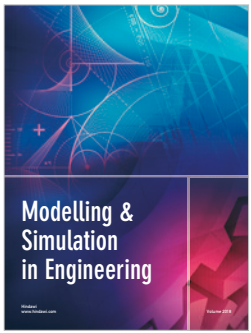

Modelling & Simulation in Engineering

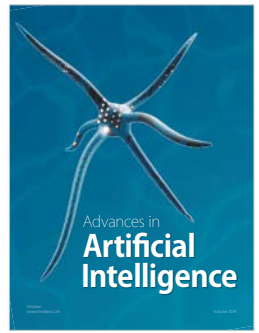

Advances in Artificial Intelligence

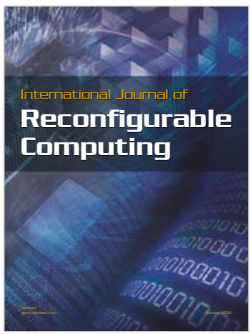

International Journal of Reconfigurable Computing

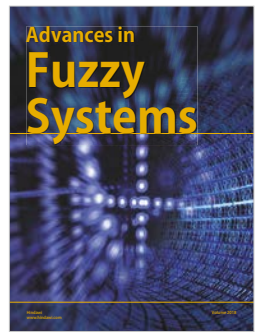

Advances in Fuzzy Systems

Hindawi

Submit your manuscripts at

www.hindawi.com

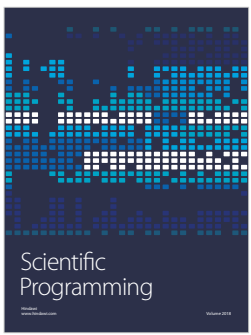

Scientific Programming

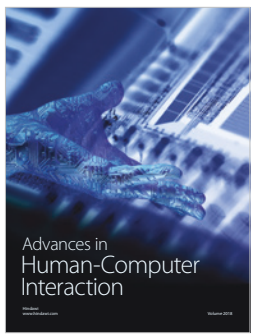

Advances in Human-Computer Interaction

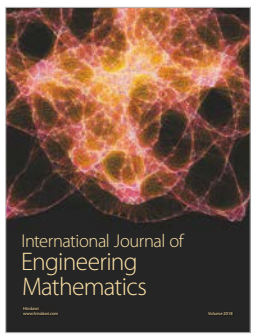

International Journal of Engineering Mathematics

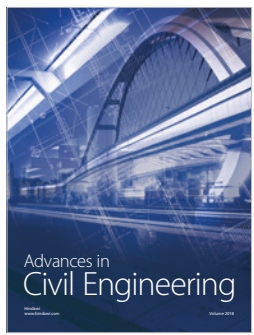

Advances in Civil Engineering

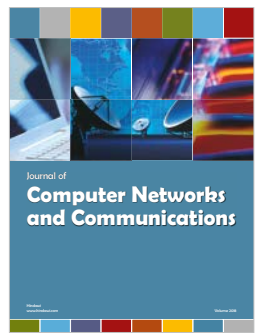

Journal of Computer Networks and Communications

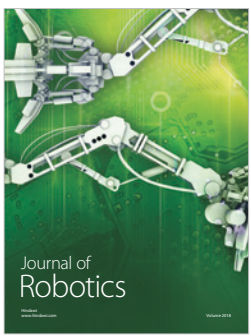

Journal of Robotics

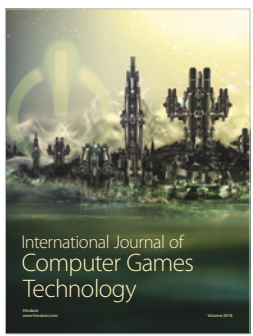

International Journal of Computer Games Technology

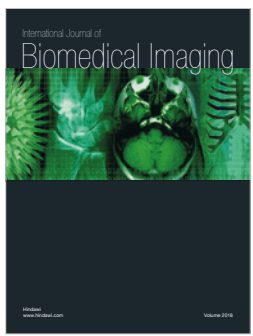

International Journal of Biomedical Imaging

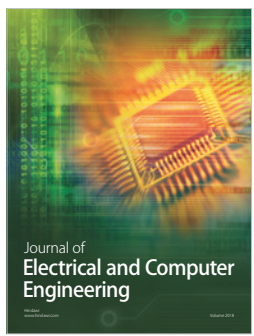

Journal of Electrical and Computer Engineering

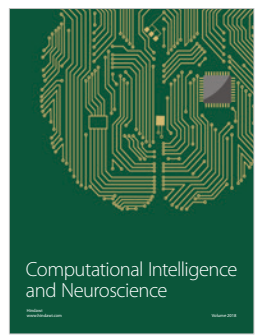

Computational Intelligence and Neuroscience